# Collaborative learning and patterns of practice

Irene-Angelica Chounta

University of Tartu, Estonia

chounta@ut.ee

## Introduction

Collaborative learning is an educational approach that involves two or more people who learn while working together on a joint task in order to achieve a common goal (for example, to solve a problem or to design a product) (Dillenbourg, 1999a). Collaborative learning builds on the idea that learning is a social activity which takes place when learners interact with their social environment (Vygotsky, 1978). It is a research field with a long history that spans multiple disciplines such as sociology, psychology and learning sciences. Early studies on collaborative learning go back to the 1920s (Allport, 1924).

Research in collaborative learning explores the way learning occurs when two or more individuals work together in order to achieve a common goal. In particular, collaborative learning studies:

- the mechanisms that are activated when people communicate and collaborate;

- the means to scaffold meaningful interactions between peers that may lead to learning; and
- the practices that facilitate collaborative learning to improve the learning outcome both from the learners' and the teachers' perspectives.

On a practical level, collaborative learning approaches are commonly adopted and illustrated in classrooms when students are asked to work in groups and deliver a common assignment.

When people collaborate in a social context, meaningful interactions may occur under certain circumstances which will trigger learning mechanisms and eventually lead to learning. However, it cannot be guaranteed that these desired interactions will indeed take place and that they will result in learning. Research has shown that social interaction between students can potentially improve learning and that students benefit more when working with peers (Crook, 1996; Fawcett & Garton, 2005; Roschelle & Teasley, 1995; Slavin, 2004; Stahl, Koschmann, & Suthers, 2006). To that end, numerous pedagogical methods have been proposed to promote and to scaffold interactions among peers with the premise that these interactions will lead to learning in a social context (Brown, Collins, & Duguid, 1989; Cobb, 1994; Valsiner & Van der Veer, 2000).

In recent years, technological advances have led to the use of computers to support collaborative learning either in classroom settings or in online learning contexts. This newly-emerged research area is known as Computer-Supported Collaborative Learning (CSCL) and it studies how computer-mediated collaboration can scaffold learning (Stahl et al., 2006). On the one hand, using technology to facilitate collaborative activities makes collaboration easier on a practical level. Students have access to common resources, they can share material with the learning community, students' activities can be logged for post-analysis purposes and, most importantly, students can participate from anywhere and at any time (Crook, 1998). On the other hand, the use of technology contributes to improving learning by scaffolding argumentation, fostering co-construction of knowledge as well as supporting coordination of peers. Additionally, collaboration contributes to shaping the students' behaviours who, in a social context, do not act as individuals but as the members of a team who work together towards a common goal (Scardamalia & Bereiter, 1991; Stahl, 2005).

In this article, an overview of the background, the research approaches and the patterns of practice in the field of collaborative learning are provided. A definition of collaborative learning and an overview of fundamental aspects that shape research and practice in this field are included. Pedagogies and learning theories that are used as foundations of the field alongside goals and objectives of collaborative learning approaches are discussed. Popular patterns of practice, exploring their application in classrooms and elaborating on the state of the art around those practices in research are outlined. A discussion about important topics, open questions and future directions are provided in conclusion.

# Background

## Fundamental aspects of collaborative learning

Collaborative learning is commonly considered part of the learning sciences. However, it is an inherently interdisciplinary field that provides a common ground for other research disciplines such as sociology, computer science and cognitive psychology (Dillenbourg et al., 1996; Hmelo-Silver, 2006). Collaborative learning studies the learning mechanisms that come into play when people collaborate to reach a common goal. It also studies the means to support people who share or build common knowledge and eventually how to improve the learning process and its outcomes. However, describing collaborative learning using one definition is challenging due to the complexity of the field (Dillenbourg, 1999b) and because it takes place in a wide context (usually wider than a classroom). It involves everyday practices of people who interact within a social arena in which learning occurs under specific circumstances (Lave & Wenger, 1991). Vygotsky discussed the role of the social context with respect to learning through his theoretical approach that identified the Zone of Proximal Development (ZPD). The ZPD can be defined as "*the distance between the actual developmental level as determined by independent problem solving and the level of potential development as determined through problem solving under adult guidance or in collaboration with more capable peers*" (Vygotsky, 1978). This definition of the ZPD highlights the importance of collaboration and social interaction when it comes to learning.

Collaboration, as an activity, is described by three dimensions that impact the way we design, implement and carry out collaborative learning activities as well as how we choose means to facilitate them. These dimensions are:

1. *Time*. Synchronous activities – that is, activities that require synchronous communication between learners – usually unfold over short period of times (for example, a few hours) and students communicate over voice channels, face-to-face when co-located, or instant messaging. Asynchronous activities – that is, activities that do not require synchronous communication between learners – can take place over longer periods of time and students usually communicate via emails or written texts.

2. *Location*. Typically, collaboration is perceived as a co-located activity. That is, people who work together are expected to physically share the same space. This is not necessarily the case in collaborative learning since the use of technology can provide means for people to collaborate efficiently that are co-located as well as people that are not. In the case of co-located collaboration - for example, face-to-face meetings - technology can provide resources or assistance to support students to interact or to coordinate. In the case of distant collaboration, technology can additionally provide a virtual, common space for people to interact and communicate and thus bridge the gap created by distance.

3. *Group size*. The size of the group in collaborative contexts (that is the number of people who work together) can vary from dyads (pairs) and small teams to classroom-size (20 to 30 students) or larger (social) groups like, for example, Massive Open Online Courses (MOOCs) or communities of practice.

Several theoretical and methodological frameworks take these dimensions into account when designing and analysing learning activities (Dillenbourg, 1999b; El Helou, 2010; Kirschner & Erkens, 2013; Pozzi, 2010). There are also approaches that focus on exploring the impact and significance of the aforementioned dimensions on collaborative learning or study specific phenomena that can be traced back and attributed to these three dimensions (Antunes, Herskovic, Ochoa, & Pino, 2012; Reimann, 2009).

Early research in collaborative learning focused on the group level: the interactions between group members, the means (technological or others) that facilitate collaboration and the effectiveness of these means on collaborative practice. Later on, the research field expanded to include work from closely-related fields such as informal learning, game-based learning, educational data-mining and teacher education.

**Theoretical Frameworks**

In principle, collaborative learning can be considered a constructivist approach that focuses on the interaction between people (Suthers, 2012). As such, it builds on the principles of *Constructivism* and *Social Constructivism*.

Constructivism (Doise, Mugny, James, Emler, & Mackie, 2013; Piaget, 2002) emphasises the way humans develop in relation to their experiences. Similarly, social constructivism (Vygotsky, 1980) focuses on how humans learn by interacting with others in a social context. Social interactions can expose people to divergent views and require them to familiarise themselves with new ideas and experiences. In a social arena, humans are expected to be able to elaborate on their own views and perceptions and to externalise knowledge. To enable these interactions and potentially turn them into learning experiences, learners have to establish a common ground which allows them to exchange information, share and build knowledge. In this sense, learning is not just the end goal (or the outcome of a collaborative activity) but it describes the whole process of working with others in a social context and towards a common goal.

The increasing popularity of the field that came along with technological advances in computer science has led to the adaptation of constructivist pedagogical approaches in order to address the specific needs of collaborative learning. Examples of popular theoretical trends in collaborative learning include the theories of *collaborative knowledge building* (Scardamalia & Bereiter, 1991), *group cognition* (Stahl, 2005), *knowledge-creating organisations* (Nonaka & Takeuchi, 1995) and *construction of shared meaning* (Koschmann, Hall, & Miyake, 2002).

**Research methodologies**

Research methodologies in collaborative learning can generally be classified as *quantitative* or *qualitative*. However, in recent years there has been an effort to move to mixed methods approaches that involve triangulation between different methodologies.

*Quantitative approaches* aim to examine the relationship (if any) between factors, to assess the impact of independent variables on dependent variables and to detect causal relationships. In collaborative learning, a typical dependent variable is the learning outcome, as assessed by pre- and post-knowledge tests or with respect to the successful outcome of a learning activity. Quantitative approaches are typically *systemic*, in the sense that they involve the use of computational models to model the relationship between quantifiable variables or features and their effect on the collaborative learning process (Greeno & Engeström, 2006). Quantitative methodologies include descriptive research, explanatory methods and experimentation (Johnson, 2001).

*Qualitative* approaches are usually *descriptive* or/and *dialogic.* Such approaches involve field observations of processes or phenomena and their unfolding over time in a specific context and they focus on the analysis of verbal (or written) interaction between peers to understand the ways people work together (Valsiner & Van der Veer, 2000). Qualitative methods used in collaborative learning research typically derive from ethnology, ethnomethodology, discourse analysis and grounded theory (Lipponen, 2002).

**Goals and objectives**

The main objective of collaborative learning is to improve the learning outcomes. This objective can be broken down into sub-goals that may contribute to improve students' performance, such as to improve the quality of collaboration, to scaffold students' motivation and to foster creativity.

Thus, research in the field of collaborative learning extends towards several subfields (Lehtinen, Hakkarainen, Lipponen, Rahikainen, & Muukkonen, 1999; Lipponen, 2002). It can, however, be grouped into two main categories:

1. Studying *new pedagogical methods* and practices that aim to improve learning.
2. Exploring, mapping and understanding the underlying *collaborative mechanisms* and the factors that affect the outcome of a collaborative activity.

The aforementioned categories are neither mutually exclusive nor independent. On the contrary, research usually intertwines both categories. Furthermore, it is important to ensure the effective integration of technological affordances and pedagogical interventions into the learning design. Therefore, there is a great need for analytical and evaluative approaches in order to inform research and practice. However, analytical and evaluative approaches as well as research regarding learning design are beyond the scope of this article.

## Research practices in collaborative learning

Collaborative learning includes a wide variety of learning activities that range from collaborative problem-solving, argumentation and project-based activities to game-based or online learning. In this section, an overview of collaborative practices that aim to scaffold learning in a classroom setting are provided. In particular, work practices that stem from the following collaborative learning approaches are presented:

- Collaborative learning over shared artifacts and representations
- Project-based learning
- Game-based learning
- Collaborative learning scripts

These work practices are not necessarily collaborative, in the sense that they can also facilitate individual learning. However, in this article, these practices are explored as collaborative approaches. Furthermore, these approaches are not necessarily exclusive, but they can be combined to accommodate different learning contexts and goals. For example, it is an established practice to combine collaborative scripts with the co-construction and use of shared representations in order to teach argumentation (Scheuer, McLaren, Weinberger, & Niebuhr, 2014).

In this section, practices that are usually applied in classrooms have been included. Other popular approaches, such as online communities and MOOCs, are not discussed here. Nonetheless, the reader should keep in mind that the purpose of this article is to inform about common collaborative practices and recent advances. It is not an exhaustive overview of all existing literature on collaborative learning practices.

**Collaborative learning over shared artifacts**

A *learning artifact* can be any object created by students during a learning activity that can provide "*evidence*" of the learning process (Kafai & Resnick, 1996). The collaborative construction of artifacts is a typical activity that can promote learning, elicit reflection and facilitate communication between peers. In this case, the learning artifact is the outcome of a collaborative process and as such it reflects the quality of collaboration as well as the potential learning gains of the collaborators (Hoppe, 2009). Collaboration over shared artifacts can facilitate collaborative knowledge building by promoting the transition from tacit knowledge (that is, internalised knowledge that is difficult to communicate to peers verbally or in written form) to explicit knowledge (that is, formal and concrete knowledge that the learner can communicate to peers) through externalisation but also to support reflection by transforming explicit to tacit knowledge through internalisation (Nonaka & Takeuchi, 1995).

Examples of learning artifacts commonly used in classroom settings are diagrammatic representations – such as algorithmic flowcharts, concept maps and argumentation diagrams – but also co-written reports, bricolages or even collaborative portfolios that represent the work of a student group over time. We will explore examples of learning artefacts through a range of research studies in the remaining paragraphs in this section.

Suthers (2005) and Suthers and Hundhausen (2003) studied the use of collaboratively-created artifacts, such as graphical representations of hypothesis and relations. They found indications that the use of shared graphical representations can positively influence collaborative learning by enabling communication and stimulating discussion and co-construction of knowledge. Analysis of collaborative learning activities showed that students interact with and through the artifact in a similar way as they would during a discussion. This interaction over a shared artifact enables argumentation and supports participants in reaching an agreement. It was also suggested that the criteria for supporting participants' grounding and knowledge co-construction were present in such a setting (Suthers, 2005).

Concept maps (that is, diagrammatic representations that map the relationships between concepts) have been widely used in collaborative learning. Research has shown that the use of concept maps as a collaborative tool promotes the externalisation of knowledge (Hoppe & Gassner, 2002) and fosters critical thinking and active learning (Daley, Durning, & Torre, 2016; Kinchin, Cabot, & Hay, 2008). Manske and Hoppe (2016) explored the use of concept maps on a meta-level, to inform students and teachers about the progress of collaborative knowledge construction and to scaffold reflection. To that end, they built structured visual representations of semantic concepts based on a model of (shared) student knowledge. The authors experimentally demonstrated how visual representations of collaborative knowledge building can support learners and teach them epistemic fluency.

Collaborative construction of argument diagrams has been shown to be beneficial, in particular, for learning to argue and to co-construct knowledge (Chounta, McLaren, & Harrell, 2017; Scheuer et al., 2014). Schwarz and De Groot ((2007) used graphic tools to promote critical reasoning for argumentation during a history class. Their findings suggest that the co-construction of shared representations benefits students with respect to coherence, decisiveness and openness.

Voyiatzaki et al. (2004) studied the case of teaching algorithms through peer collaboration over co-construction of diagrammatic representations (graphs) in secondary school classrooms. To facilitate the collaborative building of graphs, the authors used a computer-based environment adapted for the needs of 15-year-old students with minimum computer-related experience. According to the authors, the students were able to adapt to the needs imposed by the setting, they collaborated without difficulties and the use of technology did not negatively influence the students' activity. The authors reported no significant learning gains but pointed out important issues that related to resource allocation that should be taken into account when designing collaborative activities for school classrooms.

Kahrimanis et al. (2009; 2011) studied the collaborative creation of algorithmic flowcharts in the classroom for teaching a first-year university course on "Introductions to Algorithms". Their results indicated that the produced artifacts (that is, the collaboratively-created flowcharts) can capture and reflect the quality of collaboration between peers and they can be used as a post-assessment tool for teachers and researchers.

**Collaborative project-based learning**

Collaborative project-based learning (PBL) is an educational approach that aims at teaching students by engaging them in pursuing solutions to problems through investigation (Thomas, 2000). Learning activities are driven by projects that the students carry out collaboratively in teams. During the process of working together on a common project that reflects the knowledge and skill of the group, the students learn on an individual level through inquiry and self-reflection and on a group level through researching and creating artifacts along with their peers for the purpose of the project.

The outcome of a project is usually a product that addresses the learning objective (Blumenfeld et al., 1991). Typically, the students have to engage in various design, problem-solving and management tasks while they interact with their peers to successfully complete the assigned project (Thomas, 2000). The learning outcome of a project-based learning activity cannot be predetermined. Therefore, students and teachers have to continuously monitor, evaluate and improve their practice in order to achieve the desirable learning outcome (Barron et al., 1998). Thus, project-based learning is considered an innovative and promising approach for fostering important skills for the 21st century learner, such as critical-thinking, collaboration and creativity (Bell, 2010; Chounta, Manske, & Hoppe, 2017).

Schneider, Synteta and Frété (2002) adopted the idea of project-based learning for web-based educational approaches. In such approaches, the teacher neither instructs nor teaches students actively. The teacher rather encourages students to work and learn independently, while – at the same time – facilitating, monitoring and evaluating the students' practice. Project-based learning moves away from the traditional teacher-centered model that is usually adopted in education.

Project-based learning is characterised by three principles, according to Schneider, Synteta and Frété (2002): (1) learning that involves students in real-world projects through which they develop and apply skills and knowledge; (2) learning that requires students to draw information from multiple resources in order to solve problems; and (3) learning in which curricular outcomes can be identified up-front, but in which the outcomes of the student's learning processes are neither predetermined nor fully predictable. In the next paragraphs, we explore examples of project-based learning approaches.

Han, Capraro, and Capraro (2015) explored the application of project-based learning in science, technology, engineering and mathematics (STEM) curricula. They studied how project-based activities affected high-school students of different performance levels over a long period of time (3 years). The results of their study showed that low performing students benefit more from PBL activities than medium and high performers. Low performing students demonstrated significantly higher growth rates on mathematics scores than high and middle performing students over 3 years. In the same study, it was shown that a student's ethnicity and economic status were significant predictors of academic achievement. This implied that students from different backgrounds than the dominating one may have more opportunities to communicate with peers and teachers than they would in a traditional classroom and, thus, they benefit more from PBL activities with respect to performance. This finding has been confirmed by similar studies (Capraro, Capraro, Yetkiner, Rangel-Chavez, & Lewis, 2010).

Lee, Huh, and Reigeluth (2015) studied collaboration as a 21st century skill that can be acquired as a PBL outcome. Furthermore, they attempted to identify how social skills relate to conflicts and how the lack of such skills affects collaboration on individuals and groups. Their findings suggested that task and process conflicts were often transformed into relationship conflicts when students lack social skills. However, when it came to reducing conflicts within groups and to promoting collaboration, group-level social skills were more influential than individual social skills.

Avouris et al. (2010) reported and discussed their experience when using a mixed approach to teaching an introduction to programming and algorithms during a first-year university course. Their approach combined synchronous collaborative problem-solving along with asynchronous collaboration through project work.

**Collaborative game-based learning**

Game-based learning (GBL) is an educational approach that capitalises on *gameplay* in order to achieve predefined learning goals (Shaffer, Halverson, Squire, & Gee, 2005). Contrary to gamification

- where gaming features, such as rewards and badges, are used to offer incentives and to motivate students - game-based learning involves the redesign of learning activities so that they will assimilate the fundamental characteristics of games: plots, artificial conflicts and rules of play (Salen & Zimmerman, 2004).

Research argues that educational games can scaffold students' motivation and improve their performance (Burguillo, 2010; Papastergiou, 2009). Furthermore, game-based learning can offer opportunities for constructive interactions with others that may lead to successful learning episodes. In this context, Voulgari and Komis (2010, 2011) explored the use of Massively Multiplayer Online Games (MMOGs) as environments for the emergence of collaborative learning. The authors studied how inherent elements of MMOGs can facilitate collaborative learning and how it is possible to apply collaborative learning principles to the design of MMOGs to promote effective interactions among players. Next, we provide some examples of collaborative game-based learning applications.

Sung and Hwang (2013) used a collaborative game-based learning environment to support students to organise and share knowledge during gameplay. They found that the collaborative gaming environment had a positive effect on students' learning motivation and achievement.

Chen, Wang, and Yu-Hsuan (2015) compared single-player GBL to collaborative GBL. Even though they found significant improvements in learning outcomes that the authors attribute to GBL, no difference was found between the single-player and the collaborative condition.

Nonetheless, the authors suggest that collaborative game-based learning allows students to re-construct or co-construct knowledge, encourages collective problem-solving and discussion of rich descriptions of science concepts. Furthermore, they pointed out that positive group dynamics – where no conflicts between individuals exist – are necessary for an efficient student practice. Therefore, group formation is a key factor to success in collaborative game-based learning approaches.

**Collaborative learning scripts**

The goal of collaborative learning scripts (or collaboration scripts) is to structure and to scaffold interactions between peers in order to direct and coordinate collaborative practice (Dillenbourg & Jermann, 2007). In other words, they script the students' activity to invoke meaningful interactions that lead to learning and avoid harmful interactions that may disrupt fruitful collaboration. Collaboration scripts capitalise on the notion of scaffolding, that is, the support learners receive in order to successfully carry out a task that they would not be able to accomplish without help. This support can refer to *content* (for example, knowledge about a domain that is required by a learning activity but that a student does not have when starting the activity) or to the *structure* of collaboration (for example, guidelines about communicating with peers or sharing information) or even to *both* content and structuring.

According to Kollar, Fischer and Hesse (2006), collaboration scripts consist of at least five central conceptual components that can be used to formally define them. These components are: (1) the *learning objectives* that are pursued with the use of scripts; (2) the *type of activities* that the scripts direct students to engage with; (3) *sequencing*, that is the planning of activities or tasks that the student will carry out; (4) the *role distribution* among the peers that defines their responsibilities and contribution; and (5) the *type of representation* that will be used to communicate the collaboration script to students.

Dillenbourg and Hong (2008) distinguish between two categories with respect to the script's focus level, that is, the dimension or aspect of collaboration that the script aims to address: (1) *Micro-scripts*, which focus on the communication between peers. Typical examples are scripts used in argumentation learning that aim to teach peers how to facilitate argument construction (Weinberger, Stegmann, & Fischer, 2010); and (2) *Macro-scripts*, which deal with the organisation of the learning activity, such as the description of learning tasks, roles and groups. Typical examples of macro-scripts are adaptations of Jigsaw, such as Concept Grid, or Arguegraph (Dillenbourg & Jermann, 2007).

The use of scripts has been studied extensively in argumentation and discourse. Stegmann, Weinberger, and Fischer (2007) studied the use of scripts to facilitate argumentative knowledge construction. Their research suggested that scripts could have a positive impact on knowledge construction and knowledge acquisition, on argumentation as well as on the formal quality of arguments. Weinberger, Fischer, and Stegmann (2005) explored the use of computer-supported collaboration scripts to facilitate argumentative knowledge construction, either by supporting single argument construction or by supporting the construction of argumentation sequences. Their results showed that learners with scripts argued better and acquired more knowledge on argumentation than learners without scripts.

Näykki et al. (Näykki, Isohätälä, Järvelä, Pöysä-Tarhonen, & Häkkinen, 2017) studied the use of collaboration scripts from a socio-cognitive monitoring and socio-emotional monitoring perspective. In particular, the authors used collaboration scripts during a first-year course for teacher education students. The objective was to study how socio-cognitive and socio-emotional monitoring processes differentiated during more and less active script discussions. Their results showed that students used the script more in the beginning of the collaborative activity as a means of oneself's orientation to the group and the task.

Wang, Kollar and Stegmann (2017) explored the use of scripts that could adapt to students' self-regulation skills. In particular, the authors tested whether the use of scripts that the students could adapt, based on their self-perceived needs, would have an impact on their self-regulation. The results of their work suggested that students who worked with adaptable scripts engaged in more metacognitive activities than those who worked with non-adaptable scripts. Also, the use of adaptable-scripts was associated with more monitoring and reflection activities than the use of no scripts.

Collaborative scripts have been criticised for leading students to superficial interactions that do not indicate actual collaboration, when they are designed poorly. It has been argued that collaboration scripts can restrain and restrict collaboration episodes from developing naturally. Over-scripting may lead students to mimic the behaviour that the script is anticipating from them - in a way, adopting a "*gaming the system*" behaviour - without in reality engaging in the cognitive and social processes that are necessary for learning (Dillenbourg, 2002).

Vogel et al. (2017) conducted a meta-analysis of scripting in order to look closer into the benefits and potential drawbacks of collaboration scripts. Their research showed that although the use of scripts leads to a small positive effect when it comes to domain-specific knowledge, scripts have a large positive effect with respect to collaboration skills.

## Important topics, open questions and future directions

In this section, important topics for collaborative learning, open questions and future research directions are discussed. In particular, the role of technology and the role of the teacher in collaborative learning is elaborated and future directions and open challenges with respect to research and practice are discussed.

**The role of technology in collaborative learning**

The use of computer systems in education is an active research field due to the growth of popularity of computer systems and their use in everyday life. Since the 1970s and up to this day (2019), Intelligent Tutoring Systems (ITSs) aim to offer targeted and personalised tutoring by tailoring the needs of students and, thus, improving learning outcomes (Corbett, Koedinger, & Anderson, 1997). ITSs-related research deals with designing activities that adapt to students' needs and prior knowledge. For ITSs, the student is considered an individual unit who learns while one practices with learning material. Because of the lack of social interaction, ITSs were criticised since education and learning is not only about providing knowledge and mastering skills, but also about integrating students in a structured society of information and knowledge (Hawkins, Sheingold, Gearhart, & Berger, 1982).

Collaborative learning addresses this criticism since it supports groups of students learning together while collaborating to achieve a common goal. In collaborative learning, technology is used to support collaboration between learners by providing means for communication and coordination, rather than providing sophisticated personalised instruction (Stahl et al., 2006). However, apart from a useful tool, technology in collaborative learning is also crucial for shaping research as well as learning and teaching practices. On the one hand, new learning paradigms such as mobile learning – that is, the educational approach where learning is mediated by using portable devices such as smartphones or tablet computers (Sharples, Arnedillo-Sánchez, Milrad, & Vavoula, 2009) – have emerged from integrating new technologies in our common practice. On the other hand, teachers adapt the way they teach in order to accommodate changes introduced by technological advances (Jahnke & Kumar, 2014).

A typical example for the impact of technology on shaping research in collaborative learning but, most importantly, on learners' practice, is the paradigm of Massive Open Online Courses (MOOCs). Over the past few years, MOOCs have become very popular. Massachusetts Institute of Technology (MIT) and Harvard University, in a joint statement in January 2017, announced that in four years 2.4 million unique users participated in one or more MITx or HarvardX open online courses while, on average, 1,554 new, unique participants enrolled per day[1].

It is evident that technology and its impact on learning should not be ignored or downplayed. On the contrary, technology can offer valuable tools that may contribute to improving teaching practices and learning gains when successfully integrated in educational approaches and designs.

**The role of the teacher in collaborative learning**

The teacher's role in collaborative learning classrooms is very different from the teacher's role in traditional classrooms (Cohen

---

[1] Source: http://news.mit.edu/2017/mooc-study-offers-insights-into-online-learner-engagement-behavior-0112

1994). In the former case, the teacher, apart from instructing, is responsible for coordinating and supporting the collaborative groups. Furthermore, the kind of support the students need is different in collaborative scenarios from that in traditional classrooms. The students need to communicate effectively and collaborate successfully with their peers. Even though one of the premises of collaborative learning is that students learn by self-reflection and self-regulation which are triggered in collaborative learning settings, interventions by teachers remain necessary when students lack the required skills. Effective collaboration does not happen out of the blue; it requires careful planning of the learning activity and even training the students-peers by using examples of good collaborative practices (Voyiatzaki & Avouris, 2014).

A common problem that teachers face in collaborative learning classrooms is to effectively monitor the practice of working groups and to provide support when necessary (Chounta & Avouris, 2016; Van Leeuwen, Janssen, Erkens, & Brekelmans, 2015). Monitoring collaborative learning activities may refer to:

- monitoring the *quality of collaboration*, that is, ensuring that the students collaborate in a meaningful and fruitful way without conflicts or miscommunication; and
- monitoring the practice of students towards *the learning objective*, that is, ensuring that the learning goal will be achieved by identifying potential misconceptions to prevent failure.

Thus, the teacher's goal is to support students when it comes to the collaborative practice but also to the learning objective (Voyiatzaki & Avouris, 2014). Furthermore, in the case of collaborative learning activities in a classroom, the unit of analysis alternates between three levels: (1) the individual student; (2) the collaborating group; and (3) the classroom (Stahl, 2006; Webb, Nemer, & Ing, 2006). In order to successfully coordinate the learning activity, the teacher has to assess the students' performances and interactions with respect to these three levels (Voyiatzaki, Polyzos, & Avouris, 2008). The workload that teachers have when monitoring and orchestrating collaborative activities becomes even higher when students are not physically present but instead collaborate in virtual groups over the internet. Therefore, there is a prominent need for tools to support teachers in course orchestration (Chan, 2011).

**Future directions**

As previously mentioned, the rapid technological advances have impact on education in general and on collaborative learning in particular. Over the past few years, research on collaborative learning has focused on how to open up *learning communities* and make them easily accessible for everyone (for example, with MOOCs).

As the name suggests, MOOCs offer open and accessible learning for almost everyone. Every individual with access to a computer and the world-wide web (WWW) can participate in a wide range of MOOC courses depending on her/his interests and learning preferences. MOOCs have also had an impact on research in collaborative learning due to their social aspect. Learning in MOOCs can be supported through social interaction of individual learners with large communities. MOOCs use online discussion forums in order to facilitate communication, information exchange and knowledge building among the members of a community. Ongoing research in MOOCs focuses on identifying what kind of user interactions can be characterised as meaningful, thus promoting communication and, consequently, information exchange and knowledge building (Gillani, Yasseri, Eynon, & Hjorth, 2014).

Social network analysis and machine learning offer valuable analytical tools to identify the *roles* students adopt when they engage in social interactions, how these roles affect student participation in MOOC forums and students' performance (Hecking, Chounta, & Hoppe, 2017; Wise & Cui, 2018). Open questions lie in the way MOOC activities to create meaningful interactions between participants are created.  Another challenge is how to provide personalised feedback to students and guide them to adopt knowledge-contributing and knowledge-building behaviours.

In addition to collaborative learning practices, there is a strong need for analysis and evaluation techniques for learning activities. Through analysis and evaluation, it is possible to inform research and practice about their outcomes and plan further improvements. The richness of data collected by applications that facilitate learning activities and the computational approaches that are available for analysing these data offers a promising approach to improve our understanding of learning. The term *Learning Analytics* is typically used to refer to the process of data collection and analysis for understanding and assessing learning (Gašević, Dawson, & Siemens, 2015). In collaborative learning, learning analytics are used to gain insights about the underlying learning mechanisms that are activated from social interaction, to assess the learning outcome but also to provide valuable information about collaboration as well. Open questions are about the way computational models to combine data from various sources in order to assess student performance, to predict the outcome of a learning activity and to prevent student dropout or failure are used. Furthermore, ongoing research explores the potential connection between learning analytics and learning design; that is, how data-driven approaches can inform course designers and support them in adapting and addressing their specific needs.

A promising trend in collaborative learning practices adapted in classroom settings stems from the *Maker movement* and modern DIY (Do it Yourself) communities (Kuznetsov & Paulos, 2010). They aim to exploit the benefits from adopting this culture in a formal educational setting. The key point of the maker culture is the creation of artifacts through a creative process in a social arena (Sharples et al., 2014). These artifacts are more than products of an assignment. In essence, they reflect the experience of the maker: what s/he learned, what s/he communicated and what s/he shared.

The maker movement came together as a "product" of a technologically-influenced DIY community (Cavalcanti, 2013). Projects in this context are typically technology-driven and combine several disciplines such as crafting and electrical engineering. While DIY projects do not necessarily require group activities, one of the core aspects of the maker philosophy is a maker space. This refers to a real, physical space, which serves as a persistent location for idea and knowledge exchange, planning, communication and implementation. Modern maker spaces build on the idea of people getting together to exchange their knowledge and their tools. This concept also inhibits social aspects: participants usually share a common space, in some cases they share accommodation and they participate in social events. Latest trends demonstrate the usefulness of institutions such as coding camps, with some companies establishing such camps as a means for vocational training (Lewin, 2014).

Promising new topics for research and practice emerge from the integration of technology in education. The traditional classroom is undergoing change and adaptation in order to accommodate new paradigms that employ technology to scaffold learning. However, technology is not seen as a tool to promote learning on an individual level, but rather as a means to support the members of a classroom (that is, the students as well as the teachers) to establish a "*networked society*". In such a society, students work together on artifacts using technology to coordinate and monitor their activities (thus, scaffolding reflection and self-regulation) and teachers orchestrate learning activities enabling social arrangements and interactions to achieve the desirable learning outcomes.